\documentclass[twocolumn,prb,aps]{revtex4}
\usepackage{graphicx}

\newcommand{\beq}{\begin{equation}}
\newcommand{\eeq}{\end{equation}}
\newcommand{\beqa}{\begin{eqnarray}}
\newcommand{\eeqa}{\end{eqnarray}}
\newcommand{\beqan}{\begin{eqnarray*}}
\newcommand{\eeqan}{\end{eqnarray*}}
\newcommand{\non}{\nonumber \\}

\def\xhat{{\hat x}}
\def\zhat{{\hat z}}
\def\nhat{{\hat n}}
\def\rhat{{\hat r}}
\def\rbf{{\bf r}}
\def\Lsc{{\cal L}}
\def\Psc{{\cal P}}
\def\Ssc{{\cal S}}
\def\Isc{{\cal I}}
\def\Tr{{\rm Tr}}
\def\la{{\langle}}
\def\ra{{\rangle}}
\def\bcdot{{\mbox{\boldmath{$\cdot$}}}}
\def\btimes{{\mbox{\boldmath{$\times$}}}}

\begin{document}

\title{An Analytical Formulation of CPMAS}

\author{F. Marica and R. F. Snider}
\affiliation{Department of Chemistry, University of British Columbia,
Vancouver, Canada, V6T 1Z1}

\date{\today}

\begin{abstract}

        An exact expression for the cross polarization between two
spin-1/2 particles is derived from the quantum Liouville equation.
This is given in the form of two integrodifferential equations. These
can be solved exactly in the static case (no sample spinning) and a
powder average easily performed numerically.  With magic angle
spinning, the neglect of certain interference terms simplifies the
numerical calculation.  A further assumption decoupling the
calculation of the sidebands gives a very simple formula that is
capable of giving a qualitative interpretation of all experimental
observations.  Examples are given illustrating typical buildup curves
and CPMAS matching profiles.

\end{abstract}

\maketitle

\section{Introduction}

        The most frequently used technique to improve the sensitivity
of nuclei with low gyromagnetic ratios and/or low isotopic abundances
in solid-state NMR is the Hartmann-Hahn cross polarization (CP) in the
rotating frame \cite{Pines}.  With this technique, polarization can be
transferred from sensitive and often abundant nuclear spins $I$ (e.g.,
protons) to the insensitive nuclear spins $S$ (e.g., $^{13}$C,
$^{15}$N).  It is driven by the flip-flop transitions where both
dipolar coupled spins $I$ and $S$ change their magnetic quantum number
in an energy conserving process.  Since this transfer of polarization
takes time to develop, CP can be observed only if the whole spin
system has reasonably long relaxation times.  The CP spin dynamics has
already been discussed more then 25 years ago and since then has been
the subject of a variety of additional studies.  The polarization of
the abundant $I$ spins (usually with a high gyromagnetic ratio) is
spin-locked in the rotating frame by a radiofrequency (rf) field. When
a second rf field of appropriate magnitude is then simultaneously
applied to the dipolar coupled rare $S$ spins (usually with a low
gyromagnetic ratio), flip-flop processes become energy conserving and
thus allow the transfer of polarization from the $I$ spins to the
unpolarized $S$ spins.  An overview of the Cross Polarization under
Magic Angle Spinning (CPMAS) can be found in a number of places, a
partial list of which is Refs.
\onlinecite{Fyfe,Mehring,Encyc,Meier94}. The emphasis in this paper is
on an exact theoretical formulation of CPMAS and no attempt is made to
review the experimental methods and/or results that have been achieved
using them.

        Several theoretical approaches have been presented for the
description of CPMAS.  One such approach is based on there being a
spin temperature, which assumes the spins are in thermal equilibrium
in an appropriate rotating frame, the recent paper by Meier
\cite{Meier92} exemplifies how this approach can be used. Another
method, due to M\"uller et al \cite{Muller}, resulted in an empirical
fitting formula which has been extensively used by many workers.  A
third type of approach is based on the use of floquet theory
\cite{Ernst,Marks} and a comprehensive program GAMMA has been written
\cite{GAMMA} which can be used for treating spin dynamics and
numerically solving the quantum Liouville equation.  A method of
describing the spin dynamics in terms of rotations by parts of the
hamiltonian has been used by Levitt \cite{Levitt91} and Zhang et al
\cite{Zhang}. Finally, there are methods that use the Mori \cite{Mori}
approach to nonequilibrium statistical mechanics.  One method is based
on projecting onto selected parts of the hamiltonian \cite{Demco}
whereas another \cite{Engelke} uses standard approximation methods for
estimating the memory kernel.  In contrast to this variety of methods,
the approach presented here utilizes the special nature of the
hamiltonian in the doubly rotating frame, and the specific properties
of spin-1/2 systems to obtain an exact integrodifferential equation
(correctly a pair of such equations) for the cross polarization.  A
solution of these equations is complicated by any time dependence of
the dipolar coupling, while an exact solution is immediately available
for a time independent and orientation fixed coupling constant.  The
powder average of the resulting solution can be done numerically.

        While the spin temperature approach is very appealing in that
it provides a relatively simple explanation for the qualitative
behaviour of cross polarization, it has a number of weaknesses.  First
of all, it is noticed that the spin system suffers no (or at most
negligible) relaxation during a cross polarization experiment.  Thus
there is no basis for an equilibrium to be obtained and thus no basis
for assuming a thermal distribution.  An argument might be made for
ergodic behaviour but we do not find that appealing.  Second, the
thermal equilibrium is to be in the doubly rotating frame, in which
the hamiltonian must still be time dependent if magic angle spinning
is performed.  Thus a selection has to be made in some way to extract
a time independent hamiltonian in order to have a thermal
distribution.  Exactly how this is done is arbitrary. Thirdly, only
one Hartmann-Hahn matching condition can be described at a time, after
selecting out the right time independent hamiltonian. The theory
presented here is a purely dynamical theory (no relaxation is
involved), is capable of treating the magic angle spinning exactly
(approximations simplifying the calculation are made, but all time
dependence could be included), and all four peaks of the CPMAS
matching profile are obtained from the same calculation.

        The paper starts by reviewing some relevant properties of
spin-1/2 systems together with their rotational properties.  These are
used in Sec. III to examine the specific nature of the spin
hamiltonian in the doubly rotating frame and thus show that the
relevant operator space for describing cross polarization is only 6
dimensional.  Section IV then derives a formal expression for cross
polarization, described in terms of two integrodifferential equations.
Section V gives an exact expression for the static (no spinning) case
and exemplifies the purely oscillatory time dependence for a constant
dipolar coupling strength (fixed direction of the vector joining the
particles containing the spins), and also the destructive interference
effects that occur in a powder sample.  Sec. VI describes the effect
of magic angle spinning and a short discussion ends the paper. Figures
illustrating a variety of different behaviors are presented.

\section{A system of two Spin-1/2 Particles}

        The spin operators for the two particles are labelled as
$S_x,~S_y,~S_z$ and $I_x,~I_y,~I_z$.  There are also the identities
$1_S$ and $1_I$ for the respective spin spaces.  As spin-1/2
particles, these operators satisfy the product conditions
\beq\label{prodcond}
S_xS_y={i\over2}S_z,\hspace{1cm}I_xI_y={i\over2}I_z,
\eeq
\beq\label{sqcond}
S_xS_x={1\over4}1_S,\hspace{1cm}I_xI_x={1\over4}1_I,
\eeq
and their cyclic permutations.  The first set of product conditions
are of course just part of the commutation relations
\beq
[S_x,S_y]_-=iS_z\hspace{1cm}[I_x,I_y]_-=iI_z,
\eeq
and their cyclic permutations.  [Throughout this work, $\hbar$ is set
equal to 1.]  The four operators for each spin (including the
identity) implies a total operator space of 16 dimensions, which
exactly coincides with the dimension of a matrix representation of a
system of two spin-1/2 particles.  Thus any operator can be expressed
in terms of the sum of products of $S$ and $I$ spin operators.  For
simplicity of presentation, the identity operators are usually not
written explicitly.  Thus the operator $S_z$ is understood to stand
for the product $S_z1_I$, etc.

        Another important property of spin operators (not just of
spin-1/2) is that the commutator of a spin operator generates a
rotation.  Setting the commutator superoperator
\beq
\Ssc_z A\equiv[S_z,A]_-,
\eeq
the exponential superoperator
\beq
e^{-i\theta\Ssc_z}S_x=
e^{-i\theta S_z}S_xe^{i\theta S_z}=S_x\cos\theta+S_y\sin\theta
\eeq
rotates the $S$-spin operator on which it acts (here $S_x$) about the
$\zhat$ axis.  $\Isc_z$ is used to denote the corresponding commutator
superator for $I_z$.  Analogous commutator superoperators for the
other spin directions can be defined but are not needed in this work.

\section{Natural Partitioning of the Operator space in the
Doubly-Rotating Tilted Frame}

        This is phrased in the tilted frame, namely the rf fields are
labelled as being in the $\zhat$ direction and the spin operators of
the truncated (often referred to as the secular part of the) dipolar
hamiltonian lying along the $\xhat$ direction.  Thus the hamiltonian is
\cite{Levitt86}
\beq\label{hamiltonian}
H(t)=\omega_{1I}I_z+\omega_{1S}S_z+2b(t)I_xS_x.
\eeq
The angular velocities $\omega_{1I}=|\gamma_IB_{1I}|$ and
$\omega_{1S}=|\gamma_SB_{1S}|$ are proportional to the radio frequency
field amplitudes $B_{1I}$ and $B_{1S}$.  In the laboratory frame,
these rf fields are oscillating at the respective resonance Zeeman
frequencies determined by a large static magnetic field in the $\xhat$
direction.  $b(t)$ is the (possibly) time dependent heteronuclear
dipolar coupling strength.  Fundamentally this coupling strength
depends on the angle $\theta$ between the vector $\rbf$ pointing from
the nucleus having its spin labelled $I$ to the nucleus whose spin is
labelled $S$ relative to the Zeeman field direction $\xhat$, being
explicitly given by
\beq
b(\theta)=b_0(\cos^2\theta-1/3),
\eeq
where $b_0=-3\mu_0\gamma_I\gamma_S\hbar/(8\pi r^3)$ is the dipolar
coupling constant.  Ignoring any internal (rotational or vibrational)
motion of the molecules in the chemical system, a time dependence
arises only through the physical spinning of the sample about an axis
at the magic angle [$\arccos(1/\sqrt3)$] to the Zeeman field ($\xhat$
direction).  Thus $\theta$ becomes time dependent and $b(t)$ is
written as a shorthand for $b(\theta(t))$. There is inherently a
remaining orientation dependence of $b(t)$, associated with the
original orientation $\theta(0)$ of $\rbf$.  In a powder sample there
is a random distribution of such angles so an average over $\theta(0)$
needs to be made for such samples.  But for the analysis of the spin
dynamics, only the possible time dependence of $b(t)$ is important, so
only this time dependence is explicitly displayed.  Later on, when a
powder sample is considered, the extra averaging over initial
orientations (or its equivalent) is made.
 
        It is also noted that, if $b$ is time independent (and has a
fixed interspin orientation, $\theta$), the energy eigenvalues of this
hamiltonian are $\pm(1/2)\sqrt{b^2+(\omega_I\pm\omega_S)^2}$.  These
eigenvalues have been noted (at least) by Levitt, et al
\cite{Levitt86}, but their method of solving the spin dynamics, as do
some other treatments, for example \cite{Levitt91,Zhang,Wu}, is based
on a separation of the hamiltonian, Eq. (\ref{hamiltonian}), into two
commuting parts.  Neither the eigenvalues, nor their respective
eigenvectors, seem to be standardly quoted in the literature.  They
are listed for reference in Appendix A.  The method used in this paper
does not explicitly use these eigenvalue properties nor the above
noted separation of the hamiltonian into commuting parts.

        The purpose of this section is to catalog how the 16 spin
operators behave under the action of the commutator superoperator
$\Lsc(t)$ for the hamiltonian,
\beq
\Lsc(t)A\equiv[H(t),A]_-=\Lsc_{\rm rf}A+\Lsc_{\rm d}(t)A.
\eeq
This is listed as a sum of radio frequency and dipolar parts.  It is
to be shown that this set of operators is partitioned into four
dynamically separate sets of operators.  Crucial to this partitioning
is the vanishing of commutators of the form
\beqa\label{comm2}
[S_xI_x,S_yI_y]_-&=&S_x[I_x,S_yI_y]_-+[S_x,S_yI_y]_-I_x \non
&=&S_xS_y(iI_z)+iS_zI_yI_x \non &=&(i/2)S_z(iI_z)+iS_z(-i/2)I_z=0.
\eeqa
This relation is special to spin-1/2 systems since it depends on the
special product conditions, Eq. (\ref{prodcond}), of the spin-1/2
operators.

        Starting with the operator $I_z$, the action of $\Lsc(t)$ on
successively appearing operators is
\beqa\label{comm6}
\Lsc I_z&=&-2ibS_xI_y,\non
\Lsc S_xI_y&=&i\omega_{1S}S_yI_y-i\omega_{1I}S_xI_x+{i\over2}bI_z\non
\Lsc S_yI_y&=&-i\omega_{1S}S_xI_y-i\omega_{1I}S_yI_x \non
\Lsc S_xI_x&=&i\omega_{1S}S_yI_x+i\omega_{1I}S_xI_y   \non
\Lsc S_yI_x&=&-i\omega_{1S}S_xI_x+i\omega_{1I}S_yI_y+{i\over2}bS_z
\non  \Lsc S_z&=&-2ibS_yI_x.
\eeqa
The possible time dependence of $\Lsc$ and $b$ has not been indicated
in these equations.  This gives a closed set of 6 operators, namely
$I_z,S_z,S_xI_x,S_xI_y,S_yI_x$ and $S_yI_y$.  It is this set of
operators that govern the behaviour of a CPMAS experiment.  The set
of commutators (\ref{comm6}) have some special properties which are
useful for the subsequent analysis.  Specifically it is noticed that
the dipolar hamiltonian acts only to couple the single 1-spin
operators $S_z$ and $I_z$ to the product spin operators, whereas the
rf operators act only on the product spin operators.

        Another closed set consists of the identity $1_S1_I$ as its
sole member, since this operator is unchanged by $\Lsc$.  So it is a
set all by its own.  The same is true for the product operator
$S_zI_z$ since it commutes with both rf operators and, because of the
commutator type of Eq. (\ref{comm2}), it also commutes with the
truncated dipolar operator. Finally there are the remaining 8
operators, which can easily be shown to be all dynamically coupled, so
there is no further partitioning of the spin space.

        Since the 16 operators span the operator space, any operator,
and in particular the density operator $\rho$, can be expressed as a
linear combination of these operators \cite{Fano}.  For the CPMAS set
of operators, the relevant space consists only of 6 operators plus
the identity, the latter being retained in order that the density
operator be normalized, namely that the trace over both spins (a total
of four Dirac states) is consistent with $\Tr\rho=1$.  Of these
operators, only the identity has a finite trace, so $\rho$ must have
$(1/4)1_S1_I$ as one of its terms.  Recognizing the expansion
coefficients as proportional to the corresponding expectation values,
the CPMAS density operator can thus be expressed as
\beqa\label{rhoexp}
\rho&=&(1/4)1_S1_I+S_z\la S_z\ra+I_z\la I_z\ra+4S_xI_x\la S_xI_x\ra
\non &&\hspace{-.3cm}+4S_xI_y\la S_xI_y\ra+4S_yI_x\la S_yI_x\ra
+4S_yI_y\la S_yI_y\ra.
\eeqa
Thus the study of the evolution of the density operator is equivalent
to the study of the evolution of the expectation values of these
observables.

        It is appropriate to subdivide this set of operators further,
namely into the set of 1-spin operators $S_z$ and $I_z$, plus the
identity to preserve normalization, and the remaining set of 2-spin
operators.  Appropriate for this subdivision, a projection
superoperator $\Psc$ is defined that selects out the 1-spin part of
the operator space, thus
\beq
\Psc=(1/4)1_S1_I\Tr+S_z\Tr S_z+I_z\Tr I_z,
\eeq
where the trace is to be taken over both spin spaces.  Necessarily,
this superoperator is idempotent $\Psc\Psc=\Psc$, which determines the
numerical factors.  $1-\Psc$ is written for the projector onto the
remaining (2-spin) operators.  Because of the commutation relations,
it is noticed that
\beqa\label{commcp}
\Psc\Lsc\Psc&=&0, \non
(1-\Psc)\Lsc(1-\Psc)&=&\Lsc_{\rm rf}=\omega_{1S}\Ssc+\omega_{1I}\Isc,
\non
\Psc\Lsc(1-\Psc)&=&\Psc\Lsc_{\rm d}, \non
(1-\Psc)\Lsc\Psc&=&\Lsc_{\rm d}\Psc
\eeqa
separates the action of $\Lsc$ within the $\Psc$ and $1-\Psc$ spaces
to be entirely due to the rf fields (vanishing for the $\Psc$ space),
while the coupling between these two subspaces is due solely to the
dipolar hamiltonian.  The retention of one projector in each of the
subspace coupling cases serves to select which space is to be acted
upon and which is to be mapped into.  Again it is emphasized that any
time dependence of the dipolar coupling strength has no effect on this
separation of the different parts of $\Lsc$.

\section{The spin operator dynamics for Cross Polarization}

        After a  90$^\circ$ pulse to align the $I$-spin along the
$\zhat$ axis (doubly rotating tilted frame), the time evolution of the
expectation value $\langle S_z\rangle$ of the $S$-spin is monitored
experimentally. Thus the calculation is to use the value of $\langle
I_z\rangle(0)$ as the only nonzero spin expectation value as the
initial condition and to calculate the consequent time dependence of
$\langle S_z\rangle(t)$.  Since it is the expectation values of $S_z$
and $I_z$ that are of importance, it is the partition consisting of
these two operators that is relevant for the analysis of these
experiments. The projection superoperator $\Psc$ defined above carries
out this selection.

        The density operator $\rho$ for the spin system satisfies the
quantum Liouville equation (von Neumann equation)
\beq\label{vonN}
i{\partial\rho\over\partial t}=\Lsc\rho=[H,\rho]_-.
\eeq
Again $\hbar$ is set to 1.  The expectation value for any observable
$A$ is calculated as the trace $\langle A\rangle=\Tr A\rho$ whose
time dependence can be calculated according to
\beq
{\partial\langle A\rangle\over\partial t}=\Tr
A{\partial\rho\over\partial t}=-i\Tr A\Lsc\rho=i\Tr(\Lsc A)\rho,
\eeq
in the Schr\"odinger picture (first form) or Heisenberg picture (last
form).  From the commutation relations (\ref{comm6}), it is a set of 6
coupled equations that must be solved in order to predict the time
evolution of $\langle S_z\rangle(t)$.  Since only the $S_z$ and $I_z$
part of the operator space reflects the experimentally observed
quantities, a separation of the dynamics to emphasize these can be
useful.  The generalized master equation \cite{Zwanzig} approach does
this in a natural way.  This method is summarized in the following
subsection.

\subsection{Derivation of the generalized master equation}

        The derivation of the generalized master equation is reviewed
in its generality since there is no immediate simplification of the
derivation by specializing the superoperators to the present case.
Note that a time dependence of $\Lsc(t)$ is allowed for here,
which is standardly not considered.

        Define the relevant part of the density operator as
$\rho_1\equiv\Psc\rho$ and the irrelevant part as the remainder
$\rho_2\equiv(1-\Psc)\rho$.  Then these parts evolve according to
\beqa
i{\partial\rho_1(t)\over\partial t}&=&\Psc\Lsc(t)\rho(t)\!=\!\Psc\Lsc
\Psc\rho_1(t)+\Psc\Lsc(t)(1-\Psc)\rho_2(t), \non
i{\partial\rho_2(t)\over\partial t}&=&(1-\Psc)\Lsc(t)\rho(t)=
(1-\Psc)\Lsc(1-\Psc)\rho_2(t)  \non &&+(1-\Psc)\Lsc(t)\Psc\rho_1(t).
\eeqa
Here the possible time dependence of the coupling superoperators
$\Psc\Lsc(t)(1-\Psc)$ and $(1-\Psc)\Lsc(t)\Psc$ are explicitly
displayed.  Solve formally for $\rho_2(t)$,
\beqa
\rho_2(t)&=&e^{-i(1-\Psc)\Lsc(1-\Psc)t}\rho_2(0)\non
&&\hspace{-1.5cm}-i\int_0^t
e^{-i(1-\Psc)\Lsc(1-\Psc)(t-t')}(1-\Psc)\Lsc(t')\Psc\rho_1(t')dt'.
\eeqa
Note that it is assumed that there is no time dependence of
$(1-\Psc)\Lsc(1-\Psc)$, so the exponential evolution superoperator is
correct as written.  This is the case for the present problem.
Substitute this equation for $\rho_2(t)$ back into the differential
equation for $\rho_1(t)$,
\beqa\label{gme}
i{\partial\rho_1\over\partial t}&=&\Psc\Lsc\rho=\Psc\Lsc\Psc\rho_1
\non &&+\Psc\Lsc(1-\Psc)(t)e^{-i(1-\Psc)\Lsc(1-\Psc)t}\rho_2(0) \non
&&\hspace{1cm}-i\int_0^t\Psc\Lsc(t)(1-\Psc)e^{-i(1-\Psc)\Lsc
(1-\Psc)(t-t')}\non &&\hspace{2cm}\times(1-\Psc)\Lsc(t')\Psc\rho_1(t')dt'.
\eeqa
This is the generalized master equation of Zwanzig \cite{Zwanzig},
with an additional time dependence of the coupling superoperators.  It
describes the exact evolution of the relevant observables while taking
into account the role of the irrelevant observables in this evolution
by means of an integral over the time history (memory) of the relevant
observables.

\subsection{Application of the generalized master equation}

        For a cross polarization experiment involving spin-1/2
particles, the special properties, (\ref{commcp}), of the Liouville
superoperator allows all projection superoperators to be dropped in
the generalized master equation for $\rho_1(t)$.  Moreover, the
initial condition on the density operator
\beq
\rho(0)=\rho_1(0)=(1/4)1_S1_I+I_z\la I_z\ra(0)
\eeq
lies entirely in the relevant space, so the generalized master
equation reduces to
\beq\label{gmecp}
{\partial\rho_1(t)\over\partial t}=-\int_0^t\Lsc_{\rm d}(t)
e^{-i\Lsc_{\rm rf}(t-t')}\Lsc_{\rm d}(t')\rho_1(t')dt'.
\eeq
This is exact for this system!

        On taking the trace of this equation with $S_z$ and using the
expansion
\beq\label{rho1exp}
\rho_1(t)=\Psc\rho(t)=(1/4)1_S1_I+S_z\la S_z\ra(t)+I_z\la I_z\ra(t)
\eeq
for $\rho_1(t)$, the equation of change for the expectation value
$\la S_z\ra(t)$ is calculated to be
\beqa
{\partial\la S_z\ra(t)\over\partial t}&=&-\int_0^tb(t)\left[K_{SS}
(t-t')\la S_z\ra(t')\right. \non
&&\hspace{1cm}\left.+K_{SI}(t-t')\la I_z\ra(t')\right]b(t')dt'.
\eeqa
Here the (possibly time dependent) dipolar coupling strengths are
written out explicitly and it should be noted how they depend on
different times.  The memory kernel $K_{SS}(s)$ is defined as
\beqa
K_{SS}(s)&=&4\Tr[S_z,S_xI_x]_-e^{-i(\omega_{1S}\Ssc_z+\omega_{1I}
\Isc_z)s}[S_xI_x,S_z]_- \non
&=&4\Tr(i)S_yI_xe^{-i(\omega_{1S}\Ssc_z+\omega_{1I}\Isc_z)s}(-i)S_yI_x
\non &=&4\Tr
S_yI_x\left[S_y\cos(\omega_{1S}s)-S_x\sin(\omega_{1S}s)\right] \non
&&\hspace{1cm}\times\left[I_x\cos(\omega_{1I}s)+I_y\sin(\omega_{1I}s)
\right] \non
&=&\cos(\omega_{1S}s)\cos(\omega_{1I}s) \non
&\hspace{-1cm}=&\hspace{-1cm}{1\over2}\left[\cos((\omega_{1S}
+\omega_{1I})s)+\cos((\omega_{1S}-\omega_{1I})s)\right].
\eeqa
In the first term the commutator $[S_z,S_xI_x]_-$ arises by using the
invariance of the trace to interchange the order of commutation
arising from $\Tr S_z\Lsc_{\rm d}\cdots$.  The memory kernel
$K_{SI}(s)$ is calculated in a similar manner to give
\beq
K_{SI}(s)=(1/2)\left[\cos((\omega_{1S}+\omega_{1I})s)-
\cos((\omega_{1S}-\omega_{1I})s)\right].
\eeq

        The corresponding equation for $\la I_z\ra(t)$ is obtained by
interchanging $S$ and $I$ labels on the various terms.  Specifically
this gives
\beqa
{\partial\la I_z\ra(t)\over\partial t}&=&-\int_0^tb(t)\left[K_{IS}
(t-t')\la S_z\ra(t')\right. \non
&&\hspace{1cm}\left.+K_{II}(t-t')\la I_z\ra(t')\right]b(t')dt'.
\eeqa
The memory kernels
\beqa
K_{II}(s)&=&1/2\left[\cos((\omega_{1S}+\omega_{1I})s)+
\cos((\omega_{1S}-\omega_{1I})s)\right], \non
K_{IS}(s)&=&1/2\left[\cos((\omega_{1S}+\omega_{1I})s)-
\cos((\omega_{1S}-\omega_{1I})s)\right],\non
\eeqa
are of the same form as the previous pair.  As a consequence, the
equations for $\la S_z\ra$ and $\la I_z\ra$ can be simplified when
rewritten in terms of the sum and difference expectation values,
namely
\beq\label{diff}
{\partial\la S_z-I_z\ra(t)\over\partial t}\!=\!-\int_0^tb(t)b(t')
\cos(\Delta(t-t'))\la S_z-I_z\ra(t')dt'
\eeq
and
\beq\label{sum}
{\partial\la S_z\!+I_z\ra(t)\over\partial t}\!=\!-\int_0^tb(t)b(t')
\cos(\Sigma(t-t'))\la S_z\!+I_z\ra(t')dt',
\eeq
involving the difference $\Delta\equiv\omega_{1S}-\omega_{1I}$ and sum
$\Sigma\equiv\omega_{1S}+\omega_{1I}$ frequencies.

        It is thus shown that the calculation of the cross
polarization separates into two independent parts.  This completes the
part of the calculation which involves the spin operators, the
remaining part of the calculation requires solving the two
integrodifferential equations.  In order to carry this out, it is
necessary to specify the detailed time dependence of the dipolar
coupling strength $b(t)$.  The static [$b(t)$ time independent] and
magic angle spinning cases are discussed in turn.

\section{Static Cross Polarization}
\label{static}

        In the static case, the dipolar coupling strength $b$ is time
independent, but still depends on the angle $\theta$ of the vector
$\rbf$ separating the $S$ and $I$ spin particles and the direction
($\xhat$) of the static magnetic field.  In a powder sample there is a
random arrangement of this direction so an ensemble average of $\la
S_z\ra(t)$ over $\theta$ needs to be taken.  But this is to be done
after $\la S_z\ra$ has been calculated for each fixed direction.

\subsection{Fixed orientation of the dipolar coupling strength}

        The expectation of the difference of the spins $D(t)\equiv\la
S_z-I_z\ra(t)$ is considered first.  According to Eq. (\ref{diff}),
this satisfies the integrodifferential equation
\beq\label{difstatic}
{\partial D(t)\over\partial t}=-b^2\int_0^t\cos(\Delta(t-t'))D(t')dt'.
\eeq
Differentiating this equation gives
\beq
{\partial^2 D(t)\over\partial t^2}=-b^2D(t)+b^2\Delta
\int_0^t\sin(\Delta(t-t'))D(t')dt',
\eeq
and differentiating again
\beqa
{\partial^3 D(t)\over\partial t^3}&=&-b^2{\partial D(t)\over\partial t}
+b^2\Delta^2\int_0^t\cos(\Delta(t-t'))D(t')dt' \non
&=&-(b^2+\Delta^2){\partial D(t)\over\partial t}.
\eeqa
As a consequence, the time derivative of $D(t)$ must satisfy
\beq
{\partial D(t)\over\partial t}=A\cos(ft)+B\sin(ft)
\eeq
where
\beq
f=\sqrt{b^2+\Delta^2}
\eeq
is the natural oscillation frequency for $D(t)$.  From the fact that
$\partial D(t)/\partial t$ vanishes at $t=0$, as follows from Eq.
(\ref{difstatic}), the coefficient $A$ vanishes.  At $t=0$ the second
derivative of $D(t)$ equals $-b^2D(0)$, so $B$ is given by
$-b^2D(0)/f$.  Consequently the derivative of $D(t)$ is given by
\beq
{\partial D(t)\over\partial t}=-{b^2\over f}D(0)\sin(ft).
\eeq
Integrating this result and choosing the integration constant so as
to satisfy the initial condition, $D(t)$ is determined to be
\beq\label{solstatic}
D(t)={\Delta^2+b^2\cos(ft)\over b^2+\Delta^2}D(0)
\eeq
Expressed in terms of the spin expectation values, note that
$D(0)=-\la I_z\ra(0)$, this is
\beq
\la S_z-I_z\ra(t)=\left[{b^2(1-\cos(ft))\over b^2+\Delta^2}-1\right]
\la I_z\ra(0).
\eeq
An analogous calculation of the sum of the spin expectation values
implies
\beq
\la S_z+I_z\ra(t)=\left[1-{b^2(1-\cos(Ft))\over b^2+\Sigma^2}\right]
\la I_z\ra(0),
\eeq
where the frequency $F=\sqrt{b^2+\Sigma^2}$ is natural for the sum of
the spin expectation values.  Finally, the sum of these results
determines the time dependence of the $S_z$ expectation value,
experimentally referred to as the ``buildup curve'', namely
\beq\label{Sztstatic}
\la S_z\ra(t)={1\over2}b^2\left[{1-\cos(ft)\over b^2+\Delta^2}
-{1-\cos(Ft)\over b^2+\Sigma^2}\right]\la I_z\ra(0).
\eeq
Thus it is seen that, for a constant $b$ (specific orientation of the
vector connecting the $S$ and $I$ spin particles), the time dependence
of $\la S_z\ra(t)$ is governed by a difference between oscillating
terms.  Actually, since $\omega_{1S}$ and $\omega_{1I}$ are both
positive (they are the amplitudes of the radio frequency fields), the
sum frequency $\Sigma$ is larger than the magnitude of the difference
frequency $|\omega_{1S}-\omega_{1I}|$.  As a consequence, the
amplitude $b^2/(b^2+\Sigma^2)$ of the sum frequency contribution is
smaller than the amplitude of the difference frequency contribution,
implying that the rapid oscillations of the sum frequency are likely
lost as noise on top of the slower oscillations of the difference
frequency.

        Equation (\ref{Sztstatic}) has been obtained by Levitt
\cite{Levitt91} and Zhang et al \cite{Zhang} by using a sequence of
rotations using effective spin operators based on a decomposition of
the hamiltonian into two commuting parts.

        Experimentally, the matching profile is obtained by scanning
the rf amplitude of the $S$-spin channel and the (relatively) longtime
average of $\la S_z\ra(t)$ measured.  This should be compared with
\beq\label{bconst}
\overline{\la S_z\ra(t)}_{\rm time}={1\over2}\left[{b^2\over b^2+
\Delta^2}-{b^2\over b^2+\Sigma^2}\right]\la I_z\ra(0).
\eeq
This time average is dominated by the difference frequency term which,
if the dominance is sufficient, implies that the experimental plot of
$\overline{\la S_z\ra}_{\rm time}$ versus $\omega_{1S}$ is Lorentzian.

\begin{figure}
\includegraphics[scale=.4]{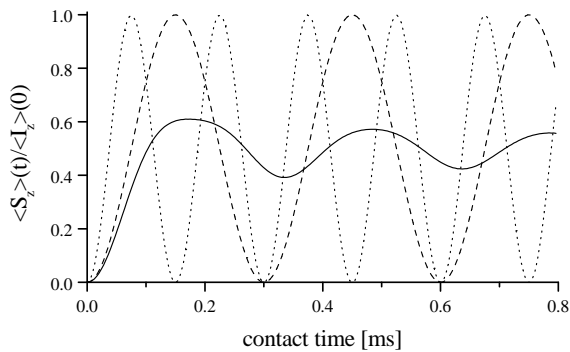}
\caption{Buildup curve for the static Hartmann-Hahn matching
condition $\omega_{1I}=\omega_{1S}=60$ kHz and a dipolar coupling
constant $b_0$ of 10 kHz. The solid line is for the powder, the dashed
line for the value of $b(\theta=\pi/2)=-10/3$ kHz, and the dotted line
for the value of $b(\theta=0)=20/3$ kHz contributing to the curve for
the powder.}
\end{figure}

\subsection{Powder average}

In a powder all orientations must be averaged over, so the cross
polarization time dependence is
\beqa\label{Szpow}
\overline{\la S_z\ra(t)}_{\rm powder}&=&{1\over4}\int_0^\pi
b^2(\theta)\left[{1-\cos(f(\theta)t)\over b^2(\theta)+\Delta^2}
\right.\non &&\hspace{-0.5cm}\left.-{1-\cos(F(\theta)t)\over b
(\theta)^2+\Sigma^2}\right]\sin\theta d\theta\la I_z\ra(0),
\eeqa
with an analogous expression for the time average.  Note that since
there is no $\phi$ dependence in $b$, the average over $\phi$ is
unnecessary while the frequencies $f$ and $F$ are $\theta$ dependent
through their dependence on $b(\theta)$.  Figure 1 shows the effects
of the angle averaging on the time dependence of $\la S_z\ra(t)$, two
of the rapidly oscillating contributions to this powder average are
also shown.  Compare the corresponding depolarization curves in Ref.
\onlinecite{Meier94}.  The corresponding matching profile, namely a
scan of the time and powder average $\overline{\la S_z\ra}_{\rm time,
powder}$ for different $\omega_{1S}$, is given in Fig. 2. A
Lorentzian, picked to have the same height and width at half height,
is also plotted.  This shows that the powder CP line shape is close to
Lorentzian, but not quite.

\begin{figure}
\includegraphics[scale=.4]{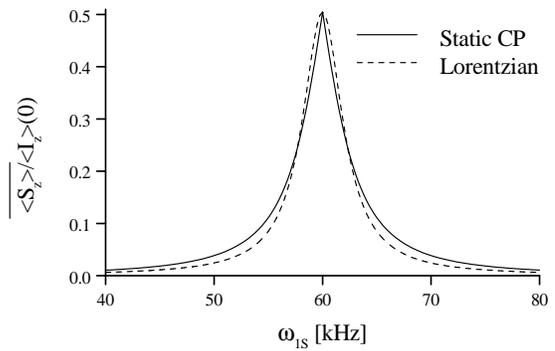}
\caption{A comparison between the Hartmann-Hahn matching
profile for a static CP of a powder sample (solid line,
$\omega_{1I}=60$ kHz and $b_0$=10 kHz) and the Lorentzian
$2.555/[5.0625+(\omega_{1S}-60)^2]$ (dashed line).  This was fitted
at the peak height and the width at 1/2 height.}
\end{figure}

\section{Cross Polarization with magic angle spinning}

        The orientation of the vector $\rbf$ connecting the positions
of the two spins is made time dependent by rotation about an axis
$\nhat$ at a frequency of $\omega_r$.  For magic angle spinning, the
angle between the rotation axis $\nhat$ and the direction $\xhat$ of
the static magnetic (Zeeman) field is $\arccos(1/\sqrt{3})$.  At that
angle the dipolar angle dependence is split into four time dependent
parts, rather than the five parts for a general splitting of the order
2 spherical harmonics.  Explicitly this time dependence is given by
\beq\label{boft}
b(t)=b_2^*e^{-2i\omega_rt}+b_1^*e^{-i\omega_rt}+b_1e^{i\omega_rt}
+b_2e^{2i\omega_rt},
\eeq
where
\beq\label{b1}
b_1={\sqrt{2}\over3}b_0\cos\theta_r\sin\theta_re^{i\phi_r}
\eeq
and
\beq\label{b2}
b_2={1\over6}b_0\sin^2\theta_re^{2i\phi_r}
\eeq
are proportional to the spherical harmonics $Y_{21}(\theta_r,\phi_r)$
and $Y_{22}(\theta_r,\phi_r)$, expressed in terms of the angles
$\theta_r,~\phi_r$ defining the direction between the particles
containing the spins with respect to the axis of rotation $\nhat$.  A
short discussion of how this splitting arises is given in Appendix B.

\subsection{General formulation for magic angle spinning}

        The dipolar coupling strength $b(t)$ enters into the
integrodifferential equations (\ref{diff}-\ref{sum}) for the cross
polarization in the form of the product $b(t)b(t')$.  From the
explicit time dependence given by Eq. (\ref{boft}), the product has a
total of 16 different terms.  Four of these are naturally combined
into two simple forms involving the time difference $t-t'$, while the
remainder have a variety of time dependences.  In detail, the
combination of time dependences is
\begin{widetext}
\beqa\label{btbt'}
b(t)b(t')&=&2|b_1|^2\cos(\omega_r(t-t'))+2|b_2|^2\cos(2\omega_r(t-t'))
\non
&&+b_2^2e^{2i\omega_r(t+t')}+b_1^2e^{i\omega_r(t+t')}
+(b_1^*)^2e^{-i\omega_r(t+t')}+(b_2^*)^2e^{-2i\omega_r(t+t')} \non
&&+b_1b_2\left[e^{i\omega_r(2t+t')}+e^{i\omega_r(t+2t')}\right]
+b_1^*b_2^*\left[e^{-i\omega_r(2t+t')}+e^{-i\omega_r(t+2t')}\right]
\non
&&+b_1b_2^*\left[e^{i\omega_r(-2t+t')}+e^{i\omega_r(t-2t')}\right]
+b_1^*b_2\left[e^{i\omega_r(2t-t')}+e^{i\omega_r(-t+2t')}\right].
\eeqa

        In the following it is assumed that only the terms involving
the time difference $t-t'$ are important, and the remainder act to
produce a variety of interference effects.  Retaining only these two
terms, the frequency difference formula (\ref{diff}) is approximated
by
\beqa\label{diffMAS}
{\partial D(t)\over\partial t}&=&-2\int_0^t\left[|b_1|^2
\cos(\omega_r(t-t'))+|b_2|^2\cos(2\omega_r(t-t'))\right]
\cos(\Delta(t-t'))D(t')dt' \non
&=&-N_{-2}(t)-N_{-1}(t)-N_1(t)-N_2(t),
\eeqa
\end{widetext}
where $D(t)$ is again used as an abbreviation for $\la S_z-I_z\ra(t)$.
The integral in this equation is naturally divided up into four terms,
whose generic definition is
\beq
N_n(t)\equiv|b_n|^2\int_0^t\cos((\Delta+n\omega_r)(t-t'))D(t')dt',
\eeq
with the obvious definition $|b_{-n}|^2=|b_n|^2$.  A solution of Eq.
(\ref{diffMAS}) is required.  This can be accomplished as in the
static CP situation by using successive derivatives of Eq.
(\ref{diffMAS}), see Sec. \ref{static}.  But it is necessary to go up
to the ninth derivative of $D(t)$ before a sufficient number of terms
are obtained in order to eliminate the $N_n(t)$'s.  This calculation
is described in Appendix C while an approximation is made in the
following subsection that provides a simple analytical estimation of
the cross polarization with magic angle spinning.

\subsection{Independent Sideband Approximation}
\label{decoupled}

        The idea here is that, near a Hartmann-Hahn matching condition,
only one of the $N_n(t)$'s is of importance, that is, of appreciable
magnitude.  As a consequence, $D(t)$ can be considered as a sum of
four terms, each of which is determined by a separate equation.  With
a similar approximation for the $\Sigma$ based $G(t)\equiv\la
S_z+I_z\ra(t)$, the total cross polarization is described as a sum of
eight terms.

        To implement this idea, $D(t)$ is written as the sum
\beq
D(t)=D_{-2}(t)+D_{-1}(t)+D_1(t)+D_2(t),
\eeq
each term of which is to be approximated as the solution of
\beq
{\partial D_n(t)\over\partial t}=
-|b_n|^2\int_0^t\cos((\Delta+n\omega_r)(t-t'))D_n(t')dt'.
\eeq
A decoupling of Eq. (\ref{diffMAS}) has thus been accomplished. Since
the equation for $D_n(t)$ has the same structure as that of Eq.
(\ref{difstatic}), so its solution is, according to Eq.
(\ref{solstatic}),
\beq
D_n(t)={(\Delta+n\omega_r)^2+|b_n|^2\cos(f_nt)\over|b_n|^2+
(\Delta+n\omega_r)^2}D(0),
\eeq
with an effective frequency given by
\beq
f_n=\sqrt{|b_n|^2+(\Delta+n\omega_r)^2}.
\eeq
On the basis that only one $D_n(t)$ has a significant contribution
when near a matching condition, it has been assumed that $D_n(0)$ is
well estimated by $D(0)$.  Adding in the corresponding approximation
for $\la S_z+I_z\ra(t)$ and doing some minor algebra, the complete
time dependence of $\la S_z\ra(t)$ is estimated by the sum of eight
terms
\beqa\label{Szdecoup}
\la S_z\ra&=&{1\over2}\sum_{n=-2,-1,1,2}|b_n|^2\left[{1-\cos(f_nt)
\over|b_n|^2+(\Delta+n\omega_r)^2}\right. \non &&\hspace{1cm}
\left.-{1-\cos(F_nt)
\over|b_n|^2+(\Sigma+n\omega_r)^2}\right]\la I_z\ra(0).
\eeqa
Here the frequencies for the sum terms are given by
\beq
F_n=\sqrt{|b_n|^2+(\Sigma+n\omega_r)^2}.
\eeq
A powder average of this result is easily calculated since the
$|b_n|^2$'s depend only on the single angle $\theta_r$.  The time
average of Eq. (\ref{Szdecoup}) is obtained by merely dropping the
cosine terms, to give an estimation of the CPMAS matching profile.

\begin{figure}[h]
\includegraphics[scale=.4]{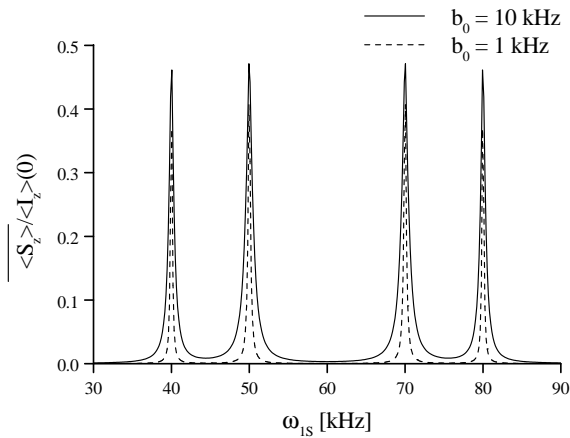}
\caption{The CPMAS matching profile for a powder sample with
$\omega_r=10$ kHz and $\omega_{1I}=60$ kHz.  The solid line is for a
dipolar coupling constant $b_0$ of 10 kHz and the dashed curve for a
dipolar coupling constant of 1 kHz.}
\end{figure}

        Figures 3 and 4 illustrate the typical matching profiles that
can be attained.  Figure 3 exemplifies what is standardly observed in
an experiment, namely four peaks at the shifted Hartmann-Hahn matching
conditions $\omega_{1S}=\omega_{1I}+n\omega_r$ for $n$=-2,-1,1 and 2.
These are due to the set of first terms in the sum of Eq.
(\ref{Szdecoup}) [with the cosine term dropped to reflect the time
average], associated with the conditions $\Delta+n\omega_r=0$.  In
such a situation the second set of terms is small because
$\Sigma+n\omega_r\gg\Delta+n\omega_r$.  But if $2\omega_r>\omega_{1I}$
then $\omega_{1I}-2\omega_r<0$ and the matching condition
$\Delta+2\omega_r$ requires a negative $\omega_{1S}$.  This is not
experimentally possible ($\omega_{1S}$ measures the strength of an rf
field) so this matching condition cannot be attained and the peak in
the matching profile is absent.  However, in this case, one of the
second set of terms in Eq. (\ref{Szdecoup}) can have a small
denominator, so a peak in the matching profile arises from that term.
This gives a negative $\la S_z\ra(t)$ so the peak is negative.  Figure
4 illustrates this case.  The structure is emphasized by choosing a
small $\omega_{1I}$ while a large $\omega_r$ is needed to allow the
matching condition $\omega_{1S}=2\omega_r-\omega_{1I}$ to be attained
for positive $\omega_{1S}$.  Meier \cite{Meier92} has a figure of
experimental data illustrating this kind of behavior, but for the
3-spin system CH$_2$.  An even more interesting profile, which does
not seem to have yet been observed experimentally, is when
$\omega_{1I}$ is even smaller, so that two matching conditions are
from the second term, namely $\Sigma-2\omega_r=0$ and
$\Sigma-\omega_r=0$, while the other two matching conditions are
$\Delta-\omega_r=0$ and $\Delta-2\omega_r$.  The result is a matching
profile with two negative and two positive peaks.  This is illustrated
by the second profile shown in Fig. 4.  It is clear that there can be
at most only two negative peaks in a matching profile.

\begin{figure}
\includegraphics[scale=.4]{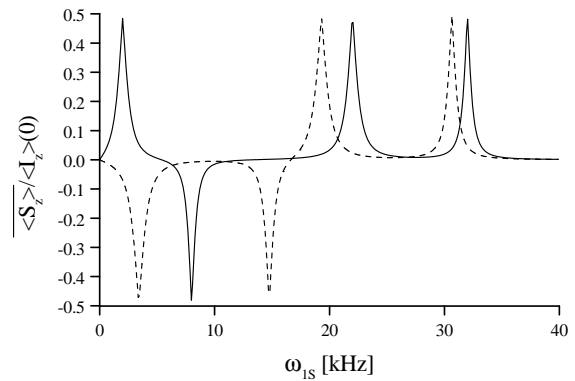}
\caption{The CPMAS matching profiles for a powder sample with a
dipolar coupling constant $b_0$ of 10 kHz.  The solid line is for
$\omega_{1I}=12$ kHz and $\omega_r=10$ kHz while the dashed line is
for $\omega_{1I}=8$ kHz and $\omega_r=11.34$ kHz.}
\end{figure}

\section{Discussion}

        The special properties of spin-1/2 systems and the special
nature of the hamiltonian, Eq. (\ref{hamiltonian}), have been used to
give a formally exact method of calculating the cross polarization
buildup curves and matching profiles for an isolated pair of spins.
For the static case (no sample spinning), this can be expressed in
closed form, Eq. (\ref{Sztstatic}), if the interspin direction is
fixed, while a powder average can be easily calculated numerically,
Eq. (\ref{Szpow}).  These results have also been obtained by Levitt
\cite{Levitt91} and Zhang et al \cite{Zhang} using a sequence of
rotations based on a particular division of the hamiltonian into
commuting parts.  It is of note that the time dependence of the cross
polarization is purely oscillatory if a single orientation of the
dipolar coupling strength is involved.  It is the sensitive dependence
of the frequency of this oscillation on the orientation, which causes
an enormous destructive interference in a powder sample.  The
decay-like nature of the envelope of the oscillations in a powder
sample, see Fig. 1, arises from this destructive interference rather
than from any spin relaxation (or spin diffusion), which of course is
not included in the hamiltonian dynamics of an isolated system.

        Magic angle spinning is formally easy to include, it is just a
case of recognizing the correct time dependence of the dipolar
coupling strengths in the integrodifferential equations (\ref{diff})
and (\ref{sum}).  But how to obtain a general solution of these
equations is not clear.  Two simplifying approximations have been
made.  First, only those terms in the product of the dipolar coupling
strengths, Eq. (\ref{btbt'}), are kept that depend on the time
difference.  It is assumed that the other time dependences lead to
minor interference effects but that has not been explored.  A method
of solving the equations with the above single approximation is given
in Appendix C, and been explored with some calculations.  The results
found, which are not presented here, indicate that the solutions are
qualitatively in agreement with those using the further approximation
of Sec. \ref{decoupled}.  This second approximation assumes that only
one term in Eq. (\ref{diffMAS}) is of importance in a range of
$\omega_{1S}$ values.  An analytic expression, Eq. (\ref{Szdecoup}),
for the cross polarization is then obtained for fixed interspin
orientation, and it is in a form in which a powder average can easily
be carried out numerically.  Its form is very simple and it is
immediately seen how the cross polarization is significant when the
standard matching conditions $\Delta+n\omega_r=0$ are met.  It is also
seen that there is another set of matching conditions,
$\Sigma+n\omega_r=0$ which can lead to an enhancement of the cross
polarization, but now with $\la S_z\ra$ and $\la I_z\ra(0)$ having
opposite signs.  There can be at most two negative peaks in the
matching profile, but there are always four physically possible
matching conditions for a given $\omega_{1I}$ and $\omega_r$.

        The method presented here, with the approximations, has the
advantage that the qualitative behavior of experimental buildup curves
and matching profiles can be given simple explanations.  The
quantitative connection between the dipolar coupling strength and the
size and width of the matching profiles, and the oscillatory nature of
the buildup curves is formally available.  How well that can be done
experimentally can also depend on the real physical system.  The spin
pair is never truly isolated, so small interactions can cause some
more interference effects, and the surroundings afford relaxation.
Such effects can surely influence the shape of the buildup curves,
contributing to a more rapid closure of the envelope of the
oscillations as time progresses.  It would also affect the height of
the matching profile peaks, but hopefully not perturb too much the
position of the peaks.

        It is interesting that the thermodynamic method can give the
same formulas for the $\Delta$-based peaks in the matching profile as
the independent sideband approximation, provided an
$\omega_r$-dependent transformation is used to get a time independent
rotating frame hamiltonian.  However, this must be done one peak at a
time whereas the present method gives all four peaks predicted by the
same formalism and there is no question about any further frame
transformation.

        Of the previously introduced methods that we are aware of for
interpreting cross polarization, the one used by Levitt
\cite{Levitt91} and Zhang et al \cite{Zhang} seems the closest to
being fundamentally based.  But it does not seem to have been extended
to include the magic angle spinning case.  It is also not clear to us
how that might be incorporated into their formalism.  We see the
formula introduced by M\"uller \cite{Muller}, and extensively used for
fitting experimental buildup curves, as a very useful fitting tool.
But since it inherently has exponential decays built into it, care
should be taken when interpreting the parameters found in terms of
physical quantities, since no relaxation is fundamental to the
explanation of the buildup curves.  We hasten to add that, as we
indicated above, in a real system the spins are not truly isolated so
relaxation effects are present and these can influence the detailed
shapes of the experimental curves, but not their fundamental behavior.

\acknowledgments

        This work was supported in part by the Natural Sciences and
Engineering Research Council of Canada.  The authors thank Professor
C. A. Fyfe and Dr. A. Lewis for pointing out a number of references
and for their support in trying to put this work into perspective. One
of the authors (F. M., on leave from the University of Bucharest,
Romania) would like to thank the Natural Sciences and Engineering
Research Council of Canada for the award of a NATO Science
Postdoctoral Fellowship held during the course of this work.

\begin{appendix}

\section{Eigenvalues and eigenvectors of $H$}

        This lists the eigenvectors and eigenvalues of the hamiltonian
defined in Eq. (\ref{hamiltonian}) for time independent $b$ and for
fixed orientation.  The basis $|mn\ra$, $m,n=\pm 1/2$ used for this
calculation is that of the simultaneous eigenvectors of $S_z$ and
$I_z$, namely
\beq
S_z|mn\ra=m|mn\ra;~~~~~I_z|mn\ra=n|mn\ra.
\eeq
As a notation similar to Zhang's \cite{Zhang}, the quantities
\beq
\omega^\Delta\equiv\sqrt{b^2+\Delta^2};~~~~~
\omega^\Sigma\equiv\sqrt{b^2+\Sigma^2}
\eeq
are introduced.  Then the eigenvectors, labelled with their
respective energy eigenvalues are:
\beqan
|\psi_{(1/2)\omega^\Delta}\ra&=&{b|1/2,-1/2\ra+(\omega^\Delta-\Delta)
|-1/2,1/2\ra\over\sqrt{b^2+(\omega^\Delta-\Delta)^2}} \\
|\psi_{-(1/2)\omega^\Delta}\ra&=&{b|-1/2,1/2\ra-(\omega^\Delta-\Delta)
|1/2,-1/2\ra\over\sqrt{b^2+(\omega^\Delta-\Delta)^2}} \\
|\psi_{(1/2)\omega^\Sigma}\ra&=&{b|1/2,1/2\ra+(\omega^\Sigma-\Sigma)
|-1/2,-1/2\ra\over\sqrt{b^2+(\omega^\Sigma-\Sigma)^2}} \\
|\psi_{-(1/2)\omega^\Sigma}\ra&=&{b|-1/2,-1/2\ra-(\omega^\Sigma-\Sigma)
|1/2,1/2\ra\over\sqrt{b^2+(\omega^\Sigma-\Sigma)^2}}.
\eeqan
It is be noted that these are NOT the same as the eigenvalues and
eigenvectors of Zhang's $H^\Delta$ and/or $H^\Sigma$, arising from
his decomposition of the hamiltonian of Eq. (\ref{hamiltonian}), even
taking into account the difference in the labelling of the coordinate
axes.

        The limits of these eigenvectors when $b\to0$ and $\Delta\to0$
or $\Sigma\to0$ give respectively the eigenvectors of the rf
hamiltonian and of the truncated dipolar hamiltonian.  In particular,
as $b\to0$ with $\Delta>0$, the expansion of $\omega^\Delta$ is
\beq
\omega^\Delta=\Delta\sqrt{1+{b^2\over\Delta^2}}=\Delta+
{b^2\over2\Delta}+\cdots,
\eeq
while $\omega^\Delta\to b$ as $\Delta\to0$.  As a consequence, the
first listed eigenvector has the limits
\begin{widetext}
\beq
|\psi_{1/2\omega^\Delta}\ra\to\cases{|1/2,-1/2\ra & as $b\to0$ for
$\Delta>0$ \cr
|-1/2,1/2\ra & as $b\to0$ for $\Delta<0$ \cr
(1/\sqrt{2})\left[|1/2,-1/2\ra+|-1/2,1/2\ra\right] & as $\Delta\to0$.
\cr}
\eeq
The limits of the other eigenvectors are obtained in an analogous
manner.

\section{Magic Angle Spinning}

        A vector based derivation is given for the effect of spinning
the orientation dependent dipolar coupling strength.

        The rotation of any unit vector $\rhat$ about an axis $\nhat$
by an angle $\chi$ produces the rotated vector
\beq
\rhat'=\nhat\nhat\bcdot\rhat+(\rhat-\nhat\nhat\bcdot\rhat)\cos\chi
+\nhat\btimes\rhat\sin\chi.
\eeq
The component of the rotated vector along the Zeeman field direction
$\xhat$ is then
\beq
\cos\theta=\xhat\bcdot\rhat'=\xhat\bcdot\nhat\nhat\bcdot\rhat+
(\xhat\bcdot\rhat-\xhat\bcdot\nhat\nhat\bcdot\rhat)\cos\chi
+\xhat\btimes\nhat\bcdot\rhat\sin\chi.
\eeq
If the unit vector $\rhat$ is expressed in a right handed coordinate
frame based on the rotation axis $\nhat$ and the $\xhat-\nhat$ plane,
\beq
\rhat=\nhat\cos\theta_r+{\xhat-\nhat\nhat\bcdot\xhat\over
\sqrt{1-(\nhat\bcdot\xhat)^2}}\sin\theta_r\cos\phi_r
+{\nhat\btimes\xhat\over\sqrt{1-(\nhat\bcdot\xhat)^2}}\sin\theta_r
\sin\phi_r,
\eeq
then
\beq
\cos\theta=\xhat\bcdot\nhat\cos\theta_r+\sqrt{1-(\nhat\bcdot\xhat)^2}
\sin\theta_r\cos(\phi_r+\chi).
\eeq
The angle dependence of the rotated dipolar coupling strength is then
expressed by
\beqa
\cos^2\theta-1/3&=&(\nhat\bcdot\xhat)^2\cos^2\theta_r-1/3
+[1-(\nhat\bcdot\xhat)^2]\sin^2\theta_r\cos^2(\phi_r+\chi) \non
&&+2\nhat\bcdot\xhat\sqrt{1-(\nhat\bcdot\xhat)^2}\cos\theta_r
\sin\theta_r\cos(\phi_r+\chi).
\eeqa
The trigonometric identity
\beq
\cos^2(\phi_r+\chi)={1\over2}[\cos2(\phi_r+\chi)+1]
\eeq
simplifies this expression.  Finally, if $\nhat\bcdot\xhat$ is chosen
as $1/\sqrt{3}$ and the angle $\chi$ is set equal to the $\omega_rt$,
the time dependence of the dipolar coupling strength is
\beqa
b_0\left(\cos^2\theta-{1\over3}\right)&=&b_0{1\over3}\sin^2\theta_r
\cos[2(\phi_r+\omega_rt)]+b_0{2\sqrt{2}\over3}\cos\theta_r\sin\theta_r
\cos(\phi_r+\omega_rt) \non
&=&b_2^*e^{-2i\omega_rt}+b_1^*e^{-i\omega_rt}+b_1e^{i\omega_rt}
+b_2e^{2i\omega_rt},
\eeqa
where $b_1$ and $b_2$ are given explicitly in Eqs. (\ref{b1}) and
(\ref{b2}).

        It is noted that the sum of the squares of the magnitudes of
these parts of the coupling strengths, when averaged over all angles,
preserves the total square of the coupling strength, namely
\beqa
\int\!\!\int b_0^2(\cos^2\theta-{1\over3})^2\sin\theta d\theta d\phi&=&
\int\!\!\int 2|b_1|^2+2|b_2|^2 \sin\theta_r d\theta_r d\phi_r \non
={8\pi\over45}b_0^2&=&{16\pi\over135}b_0^2+{8\pi\over135}b_0^2.
\eeqa
Thus the contributions of the $b_1$ terms are twice those of the $b_2$
terms.
\end{widetext}

\section{Complete Sideband Calculation}

        This appendix gives a method for carrying out the exact
calculation of $D(t)$ from Eq. (\ref{diffMAS}).  For this purpose,
the second derivative of $N_n(t)$ is
\beq
{\partial^2N_n(t)\over\partial t^2}=|b_n|^2
{\partial D(t)\over\partial t}-(\Delta+n\omega_n)^2N_n(t).
\eeq
Then the third derivative of $D(t)$ can be expressed in terms of the
first derivative and the $N_n(t)$'s, that is
\beq\label{diff3}
{\partial^3D(t)\over\partial t^3}=-\sum_n|b_n|^2
{\partial D(t)\over\partial t}+\sum_n(\Delta+n\omega_n)^2N_n(t).
\eeq
Continuing in this way, the higher odd-ordered derivatives of $D(t)$
can be calculated in terms of lower ordered derivatives and the
$N_n(t)$'s.  The first 4 equations obtained in this way (up to the
7\underbar{th} order derivative) can be used to solve for the
$N_n(t)$'s in terms of the derivatives of $D(t)$.  Substituting these
into the equation for the 9\underbar{th} order derivative produces a
differential equation for $D(t)$ of 9\underbar{th} order, but with
constant coefficients.  Such an equation has 9 solutions of the form
$D(t)=e^{\lambda t}$, whose proper combination gives the time
dependence of $D(t)$.

        To simplify this calculation, it is convenient to introduce
the notation $D^{(m)}(t)$ for the $m$\underbar{th} derivative of
$D(t)$, and the sums
\beq
SS_m\equiv\sum_n(\Delta+n\omega_r)^m|b_n|^2.
\eeq
Then the needed equations for the higher odd-ordered derivatives of
$D(t)$ are
\beqa\label{diff5}
D^{(5)}(t)&=&-SS_0D^{(3)}+SS_2D^{(1)}-\sum_n(\Delta+n\omega)^4N_n(t)
\non
D^{(7)}(t)&=&-SS_0D^{(5)}+SS_2D^{(3)}-SS_4D^{(1)}\non &&\hspace{1cm}+
\sum_n(\Delta+n\omega)^6N_n(t) \non
D^{(9)}(t)&=&-SS_0D^{(7)}+SS_2D^{(5)}-SS_4D^{(3)}\non &&
+SS_6D^{(1)}-\sum_n(\Delta+n\omega)^8N_n(t).
\eeqa
A formal inversion of the matrix of frequency powers multiplying the
$N_n(t)$'s from the 4 odd-ordered derivatives $D^{(1)}(t)$ to
$D^{(7)}(t)$ allows the $N_n(t)$'s to be expressed in terms of the
$D^{(m)}(t)$'s. Substitution into the $D^{(9)}(t)$ equation gives the
desired differential equation for $D(t)$.  This involves only
odd-ordered derivatives of $D(t)$.  If $e^{\lambda t}$ is substituted
in for $D(t)$, this gives a fourth order polynomial in $\lambda^2$
times a factor of $\lambda$.  The polynomial has 4 roots which must be
negative since the spin dynamics is governed by the quantum Liouville
equation (and thus the spin system must oscillate without any
relaxation), and there is also the zero root from the multiplicative
factor of $\lambda$.  The 4 eigenfrequencies $\Omega_m$ (square roots
of minus the polynomial roots) and the zero frequency determine the
time dependence of $D(t)$.  A further simplification arises from
noticing that $N_n(0)=0$, which implies that all odd derivatives of
$D(t)$ vanish at $t=0$.  Thus $D(t)$ must be an even function of $t$
and can be expressed as the sum
\beq
D(t)=A_0+\sum_{m=1}^4A_m\cos(\Omega_mt).
\eeq
The expansion coefficients $A_m$ ($m=0\cdots4$) still need to be
determined.

        The coefficients $A_m$ are determined from the even-ordered
derivatives $D^{(2\ell)}(t)$ evaluated at $t=0$.  These can be found
directly by differentiating Eqs. (\ref{diffMAS}), (\ref{diff3}) and
(\ref{diff5}), and evaluating them at $t=0$.  Specifically, the 5
required relations are
\beqa
A_0+\sum_mA_m&=&D(0) \non
-\sum_m\Omega_m^2A_m&=&D^{(2)}(0)=-SS_0D(0) \non
\sum_m\Omega_m^4A_m&=&D^{(4)}(0)=(SS_0^2+SS_2)D(0) \non
-\sum_m\Omega_m^6A_m&=&D^{(6)}(0)=-[SS_0(SS_0^2+2SS_2)\non
&&\hspace{1cm}+SS_4]D(0) \non
\sum_m\Omega_m^8A_m&=&D^{(8)}(0)=[SS_0^2(SS_0^2+3SS_2)\non
&&\hspace{-0.5cm}+2SS_0SS_4+SS_2^2+SS_6]D(0).
\eeqa
These equations are to be solved for the expansion coefficients.

        This determines the difference $\la S_z-I_z\ra(t)=D(t)$ of
the expectation values of the two spin operators.  A calculation of
the sum $\la S_z+I_z\ra(t)=G(t)$ involves the same treatment, but
with the sum frequency $\Sigma$ replacing the difference frequency
$\Delta$ in all the above formulas.  The cross polarization is given
by one half the sum of these two quantities
\beq
\la S_z\ra(t)={1\over2}[D(t)+G(t)].
\eeq
Note that $G(0)=-D(0)=\la I_z\ra(0)$.  To calculate the powder average
CPMAS, the above computations need to be carried out for every
possible orientation of the vector pointing from one spin to the
other.  A bonus of the approximation of retaining only the terms
$|b_1|^2$ and $|b_2|^2$ in the product of the time dependent coupling
strengths, Eq. (\ref{btbt'}), is that these quantities are independent
of the angle $\phi_r$ in the description of this orientation, so only
the $\theta_r$ average needs to be done.  Thus the powder CPMAS is
given by
\beq
\overline{\la S_z\ra}_{\rm powder}(t)={1\over4}\int_{-1}^1[D(t)+G(t)]
d\cos\theta_r.
\eeq

\end{appendix}

\end{document}